\documentclass[conference]{IEEEtran}
\IEEEoverridecommandlockouts

\usepackage{multirow}
\usepackage{multicol}
\usepackage{lipsum}
\usepackage{graphicx}
\usepackage{graphics}
\usepackage{color, colortbl}
\usepackage{array}
\usepackage{amsmath}
\usepackage{amsbsy}
\usepackage{blindtext}
\usepackage{tcolorbox}
\usepackage{amssymb}
\usepackage{scalerel}
\usepackage{mathabx}
\usepackage{verbatim}
\usepackage{booktabs}
\usepackage{rotating,tabularx}
\usepackage{mathrsfs} 
\usepackage{url}
\usepackage{adjustbox}
\usepackage{soul}
\usepackage{xcolor}
\usepackage{cite}
\usepackage{tikz}
\usepackage{algorithm}
\usepackage{algpseudocode}
\usepackage{color, colortbl}
\usepackage[first=0,last=9]{lcg}
\definecolor{LightGray}{rgb}{0.7,0.7,0.7}
\usepackage[wby]{callouts}
\usetikzlibrary{shapes.multipart}
\usepackage{array}
\usepackage{makecell}

\renewcommand{\baselinestretch}{.941}
\allowdisplaybreaks
\usepackage{amsthm}
\theoremstyle{definition}

\theoremstyle{remark}

\usepackage[utf8]{inputenc}
\usepackage[english]{babel}

\usepackage[caption=false,font=footnotesize]{subfig}
\usepackage[T1]{fontenc}
\usepackage{cite}

\begin{document}

\title{
%
Real-time 
Power System Simulation 
with Hardware Devices
through DNP3 in 
Cyber-Physical Testbed
}

\author{
\IEEEauthorblockN{Hao Huang}
\IEEEauthorblockA{
\textit{Texas A\&M University}\\
hao\_huang@tamu.edu}
\and
\IEEEauthorblockN{C. Matthew Davis}
\IEEEauthorblockA{
\textit{PowerWorld Corporation}\\
matt@powerworld.com}
\and
\IEEEauthorblockN{Katherine R. Davis}
\IEEEauthorblockA{
\textit{Texas A\&M University}\\
katedavis@tamu.edu}

}

\IEEEoverridecommandlockouts
\IEEEpubid{\makebox[\columnwidth]{978-1-7281-8612-2/21/\$31.00~\copyright2021 IEEE\hfill} \hspace{\columnsep}\makebox[\columnwidth]{ }}

\maketitle

\begin{abstract}
Modern power grids are dependent on communication systems for data collection, visualization, and control. Distributed Network Protocol 3 (DNP3) is commonly used in supervisory control and data acquisition (SCADA) systems in power systems to allow control system software and hardware to communicate. To study the dependencies between communication network security, power system data collection, and industrial hardware, it is important to enable communication capabilities with real-time power system simulation. In this paper, we present the integration of new functionality of a power systems dynamic simulation package into our cyber-physical power system testbed that supports real-time power system data transfer using DNP3, demonstrated with an industrial real-time automation controller (RTAC). The usage and configuration 
of DNP3 with real-world equipment in to achieve power system monitoring and control of a large-scale synthetic electric grid via this DNP3 communication is presented. Then, an exemplar of DNP3 data collection and control is achieved in software and hardware using the 2000-bus Texas synthetic grid.

\begin{IEEEkeywords}
DNP3 Protocol, SCADA, Hardware-in-the-Loop, Interactive Control, Cyber Security
\end{IEEEkeywords}

\end{abstract}

\section{Introduction}
Electric power systems are some of the largest industrial control systems (ICS). In these systems, operations taken by physical actuators depend on data, where this data may be delivered through a communications infrastructure. A power system is also a critical infrastructure; hence, its reliability and resilience are its key requirements. For example, it is important to ensure the integrity of generator dispatch to achieve effective utilization of energy resources and reasonable electricity prices. To achieve such goals, a reliable and secure communication network is essential. However, increasing cyberattacks are occuring worldwide~\cite{case2016analysis,knake2017cyberattack}, and more studies are showing vulnerabilities in current communication protocols \cite{xu2017review }. Thus, how to analyze, detect, and respond to cyber attacks is a vital research topic in power systems. 

Distributed Network Protocol 3 (DNP3)~\cite{IEEE1815-DNP3} is commonly used in ICS for data acquisition and control \cite{clarke2004practical}. However, several studies show the vulnerability of DNP3 and implement different types of attacks, such as event buffer flooding \cite{jin2011event}, man-in-the-middle \cite{yang2012man}, packet sniffing and modification \cite{lee2014simulated}, etc. Such explorations historically stay within the scope of only the communication network and its emulation, while neglecting real hardware devices and analysis of power system impact. To study the cyber-physical security of power systems, it is important to consider both cyber and physical elements. Recently, several hardware-in-the-loop testbeds are built with Real-Time Digital Simulator (RTDS) or OPAL-RT for power system cyber-physical security studies and algorithm validation \cite{hahn2013cyber, liu2015analyzing, poudel2017real}. Even though the incorporation of these commercial products for hardware-in-the-loop testbeds can replicate certain impacts of cyber adversaries in power systems, they cannot capture the detailed cyber attack process in the cyber network. Therefore, it is necessary to have a stand-alone power system real-time simulation that also has the functionality to communicate over a cyber network.

This paper introduces usage of a new DNP3 functionality of PowerWorld Dynamic Studio (PWDS) that allows the communication of PWDS with DNP3 clients, which enables the detailed analysis of DNP3 communication among real-time power system simulation, cyber adversaries, and industrial intelligent electronic devices (IEDs). PWDS provides an interactive simulation environment for real-time power system analysis. It can run either stand-alone or as a server; as a server, one of its capabilities is to generate IEEE C37.118 phasor measurement unit (PMU) data \cite{IEEEC37-118, overbye2017interactive, overbye2019interactive} which has been utilized in \cite{mao2020w4ips, becejac2020prime,idehen2020electric} for real-time power system data visualization, interactive control through a web interface, and digital PMU data to analog signal conversion. The addition of DNP3 functionality allows PWDS to run as a DNP3 server, generate DNP3 packets, and deliver the packets over the communication network to DNP3 clients/masters.

The U.S. Department of Energy (DOE) has funded several projects for power system cyber-physical security. In \cite{cypres}, the Cyber Physical Resilient Energy Systems (CYPRES) project is developing a secure cyber-physical modeling foundation that is truly cyber-physical: a secure end-to-end system for managing the energy system, communications, security, and modeling and analytics. PWDS with DNP3 communication capability is used in creating the hardware-in-the-loop testbed with power system modeling and analysis. This testbed is performing hardware integration over physical and emulated utility communication networks, enabling realistic security studies bridging both cyber and physical domains for CYPRES.

The main contributions of this paper are as follows:
\begin{enumerate}
\item
This paper presents a cyber-physical testbed implementation of new functionality of PWDS that enables the communication between real-time power system simulation with industry hardware devices through DNP3.

\item
We utilize an industrial control and automation device, SEL Real-Time Automation Controller (RTAC), to communicate with PWDS through DNP3.

\item 
With the synthetic Texas power grid\cite{synthetic_grids}, we present an exemplar of how to use RTAC and PWDS to mimic real-world applications of reading measurements and controlling devices using DNP3. 
\end{enumerate}



\section{PowerWorld Dynamic Studio DNP3 Functionality}
\label{PWDS}

PowerWorld Dynamic Studio (PWDS) is a transient stability based simulation running as a server. It is capable of sending and receiving data from connected clients, thus allowing multiple users to interact with the transient stability simulation. The PWDS "speaks" several protocols. It is capable of communicating via a proprietary protocol called the DS protocol (PWDSP), the IEEE C37.118 protocol \cite{IEEEC37-118} as output, and DNP3 \cite{IEEE1815-DNP3}. The use of standard protocols allows the DS to function as a stand-in for a real power system in a wide range of applications including those that are modeling cyber infrastructure. 

\begin{figure}[t]
\centering
\includegraphics[trim={0mm 0mm 0mm 8mm}, clip,width=\linewidth]{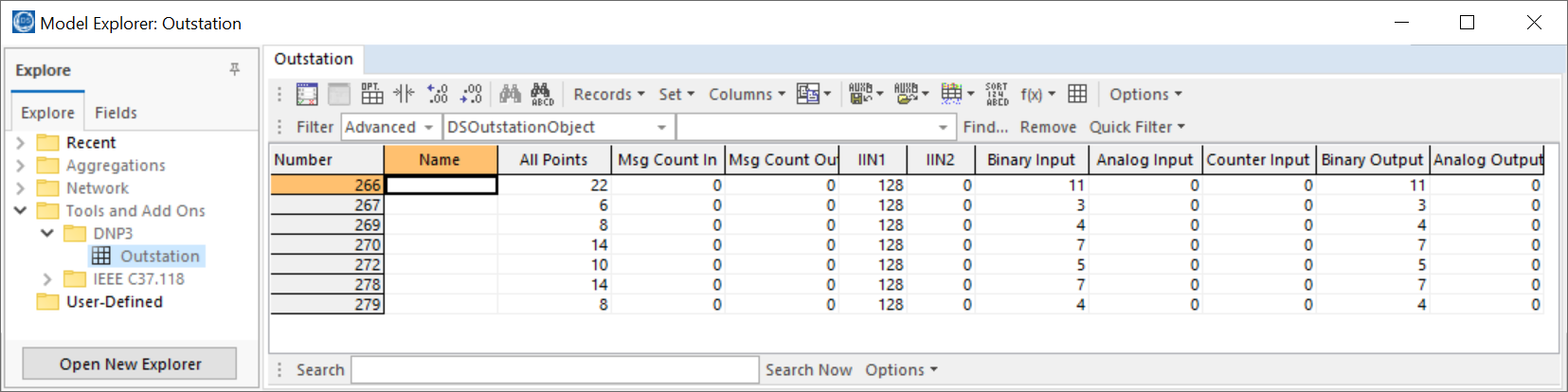}
  \caption{Outstation Records} 
  \label{figure:OutstationList}
\end{figure}

\begin{figure}[t]
\centering
\includegraphics[trim={0mm 0mm 2mm 8mm}, clip,width=\linewidth]{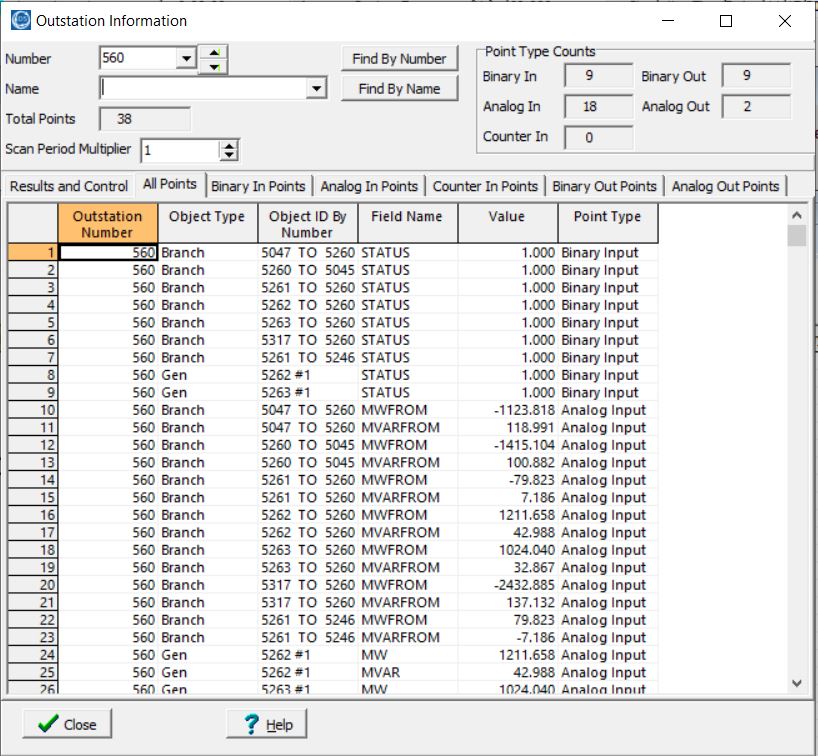}
  \caption{Outstation Information Dialog} 
  \label{figure:OutstationInformation}
\end{figure}

Just like the cases used for running the simulations, where the transient stability data must be setup ahead of time for the PWDS, a case's DNP3 data is also set up in PowerWorld Simulator before being used in PWDS. The DNP3 configurations are presented to the user in terms of two objects in Simulator: \textit{DNP3Objects} and \textit{Outstations}. The \textit{Outstation} is a container object that groups together several \textit{DNP3Objects}. A simple example of the use of an \textit{Outstation} is to group together all the points from a particular substation. However, there is no restriction in the software about which points can be assigned to an outstation, so within each DNP3 outstation, we can insert the \textit{DNP3Object} for different devices. In the PWDS, Figure~\ref{figure:OutstationList} shows the list of outstations in a sample case, while Figure ~\ref{figure:OutstationInformation} shows the dialog for an outstation.  
Dialogs in Simulator allow the user to create \textit{Outstation} objects and insert \textit{DNP3Objects}.

\begin{figure}[t]
\centering
\includegraphics[trim={0mm 0mm 0mm 8mm}, clip,width=\linewidth]{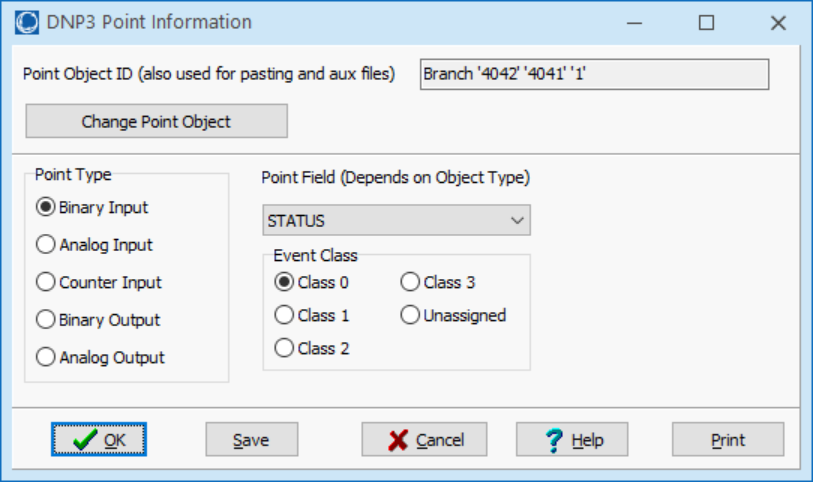}
  \caption{DNP3 Point Information Dialog} 
  \label{figure:DNP3Object}
\end{figure}

The \textit{DNP3Object} is configured using the "DNP3 Point Information" dialog as shown in Figure \ref{figure:DNP3Object}. This dialog allows the user to map an object and field in the power system model to a DNP3 point. As shown in Figure \ref{figure:DNP3Object}, there are 5 DNP3 \textit{Point Type} to choose, which are \textit{Binary Input}, \textit{Analog Input}, \textit{Counter Input}, \textit{Binary Output}, and \textit{Analog Output}. The \textit{Point Field} determines the specific data. For example, generator's \textit{STATUS} is set in \textit{Binary Input}. When the generator is on, the binary input data is represented as 1, otherwise, it is 0. In \textit{Binary Output}, the generator's status can be controlled by connected DNP3 client/master. For a generator's power data, such as its real power and reactive power output, these are set in \textit{Analog Input} as \textit{MW} and \textit{MVar}. For \textit{Analog Output}, PWDS DNP3 only supports \textit{MWSETPOINT} and \textit{VPUSETPOINT} for generators to set generator's real power and voltage values. Other devices, including loads, shunts, branchs, and buses, can be configured in the same way. 
The \textit{Event Class} determines when the data should be reported to DNP3 client, and these are customized by the user or application. Events are each placed in one of three buffers, associated with "Classes" 1, 2 and 3. In addition to these, Class 0 is defined as "static" or 
and gives the current status of the monitored data \cite{IEEE1815-DNP3}.

\section{SEL Real-Time Automation Controller (RTAC)}
\label{RTAC}

The SEL RTAC is an industrial automation and control device that supports various communication protocols, such as DNP3, Modbus, IEC 61850, etc. \cite{sheet20193555}. RTACs have been utilized in SCADA systems as remote terminal units (RTUs) for data collection and protocol conversion. The built-in IEC 61131 engine enables flexible customer-designed logic with incoming power system data and the RTAC’s system tags for substation control. The RTAC is configured through the SEL AcSELerator RTAC software (SEL-5033) that provides a programmable interface for users to configure the communication protocol types and parameters, the connection type, and user defined logic \cite{acselerator2017sel}. The embedded flex parse messaging within the SEL protocol allows users to create customized regular expressions to collect specific information, such as connected device configurations, energy measurements, etc. \cite{huang2018extracting}. Additionally, the RTAC can be accessed through its web interface, where we can configure its ethernet ports’ IP address, check connected IEDs, access the system alarms and event logs, and get a diagnostic report.


RTAC has been utilized in various power testbeds for cyber-physical security studies \cite{ oyewumi2019isaac, albunashee2019testbed}, algorithm validation \cite{shariatzadeh2014real, leonard2015real, watson2017comparing}, and data collection, conversion and control \cite{ watson2015need, becejac2020prime}. Within those applications, the RTAC is ether connected to relays working as a RTU or communicating with phasor data concentrators (PDC) to collect PMU data for real-time automated control. Any application or devices that support DNP3 communication can communicate with RTAC through serial or TCP/IP communication, which can be utilized in power system cyber-physical security studies. 

Hence, PWDS can generate DNP3 packets based on the pre-defined outstation and DNP3 tags and send them through TCP/IP network to its destination. In this way, PWDS can communicate with the RTAC and supply each outstation's DNP3 data, including the measurements, such as current, voltage, power flow, etc., and the on/off status of generators, branches, loads and shunts. This functionality provides for new approaches to study cyber-physical security among power system real-time simulation, hardware devices, and communication network.

\section{DNP3 Communication between PowerWorld DS and RTAC}
\label{Communication}

\begin{figure}[t]
\centering
\includegraphics[trim={0mm 0mm 0mm 0mm}, clip,width=0.9\linewidth]{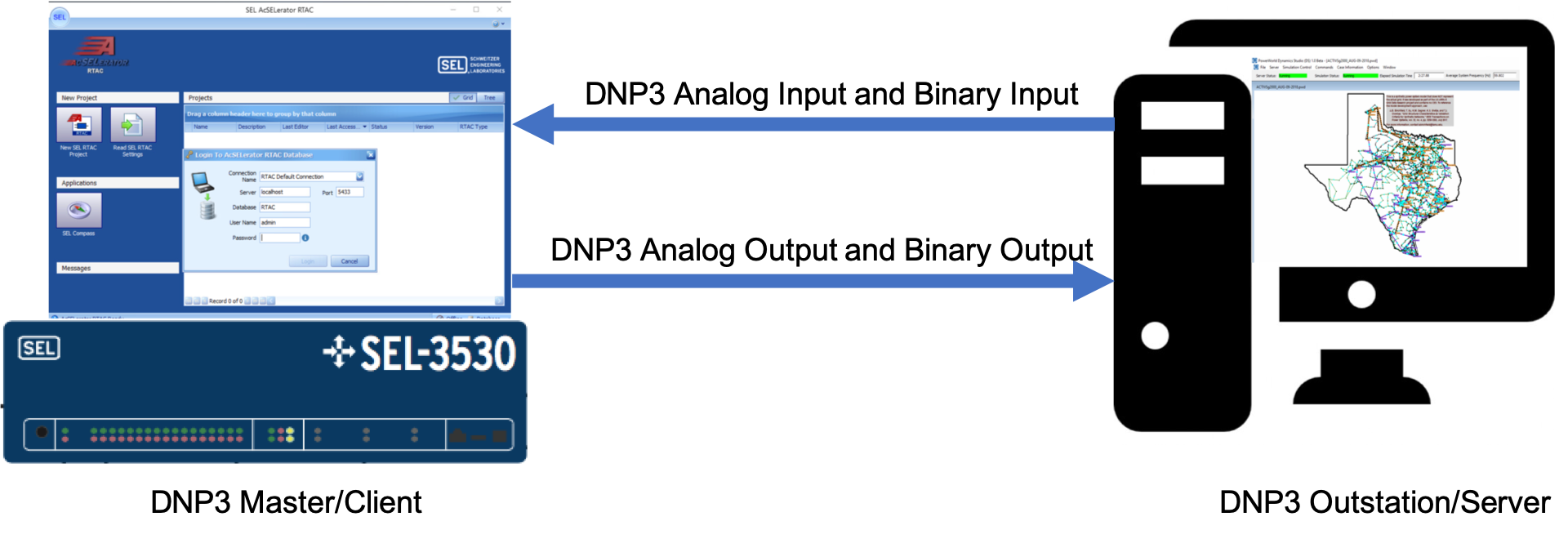}
  \caption{PWDS Communicates with SEL RTAC Over DNP3}
  \label{communication}
\end{figure}

DNP3 is frequently used in power system supervisory control and data acquisition (SCADA) systems to collect data and send control commands. As PWDS runs the simulation in real-time, each device modeled in the case has its own data including status (open/close) and measurements (e.g., real power, reactive power, voltage). With the DNP3 functionality in PWDS, the real-time simulation data is wrapped in DNP3 packets and delivered to DNP3 clients/masters. This allows the integration of real-time power system simulation with other software and hardware to replicate realistic SCADA systems with both cyber and physical elements. 

The integration of PWDS and RTAC presents one characteristic
of a cyber-physical hardware-in-the-loop testbed. The data generated by the simulation represents the field device measurements. 
The RTAC collects the data through DNP3, emulating real data transmission in the communication network. The DNP3 packets can then be captured by network analysis tools such as WireShark for further analysis. Then, within RTAC, as the DNP3 client, we can observe the collected data and control devices 
to mimic real-world operation.

\begin{figure}[t]
\centering
\includegraphics[trim={0mm 0mm 0mm 0mm}, clip,width=\linewidth]{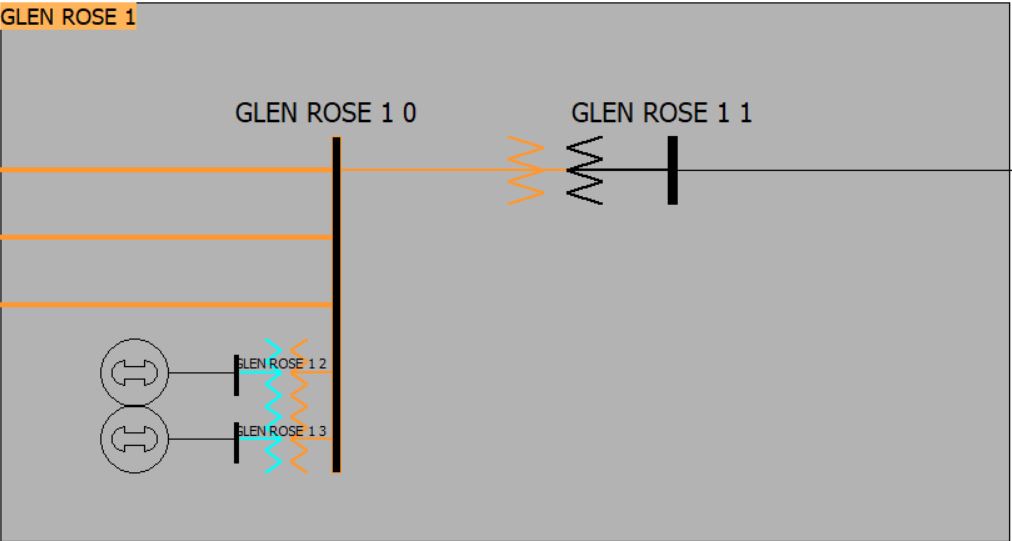}
  \caption{One-Line Diagram of Substation GLEN ROSE1}
  \label{substation}
\end{figure}

This section presents how to set the PowerWorld case to generate DNP3 packets and configure the RTAC to collect the corresponding DNP3 data. In this paper, we utilize the synthetic 2000-bus Texas case \cite{synthetic_grids} to configure the PowerWorld case and RTAC to establish the DNP3 communication and collect data and control devices. With 1250 substations in the case, we use the Substation 560 (GLEN ROSE1) as the example to show the procedure, whose one-line diagram is shown in Figure \ref{substation} with two generators, three transformers, four transmission lines and four buses. The procedure can be replicated for all other substations and devices. 

\subsection{Configuration of PWDS and RTAC}
\label{configuration}

To enable the PWDS DNP3 functionality, the first step is to configure the corresponding power system case in Simulator under the \textbf{DNP3} folder. Within the \textit{Outstation}, we can insert as many DNP3 outstations as needed. Then, in the \textit{DNP3Object}, we can insert different DNP3 \textit{Point Type}, including \textit{Analog Input}, \textit{Analog Output}, \textit{Binary Input} and \textit{Binary Output}, for various devices under corresponding \textit{Outstation}. With 1250 substations, for convenience, we configure the \textit{outstation number} based on the substation ID. Then, the devices within each substation, including generators, branches, loads, shunts and buses, and their corresponding data are configured to different \textit{Point Field} as discussed in Section \ref{PWDS}. 
Once the DNP3 configuration for the power system case is done, we can load the case to PWDS, run the real-time simulation and turn on the server. The host machine of PWDS can generate DNP3 packets at all its Ethernet ports, whose IP addresses are the DNP3 Server IP Address for DNP3 client/master to collect data for different local area networks (LANs). Regarding to DNP3 Protocol Port, it can be configured in PWDS, which is set to 20000 in this case.  

\begin{figure}[t]
\centering
\includegraphics[trim={0mm 0mm 0mm 0mm}, clip, height=2.3 in,width=3.5 in]{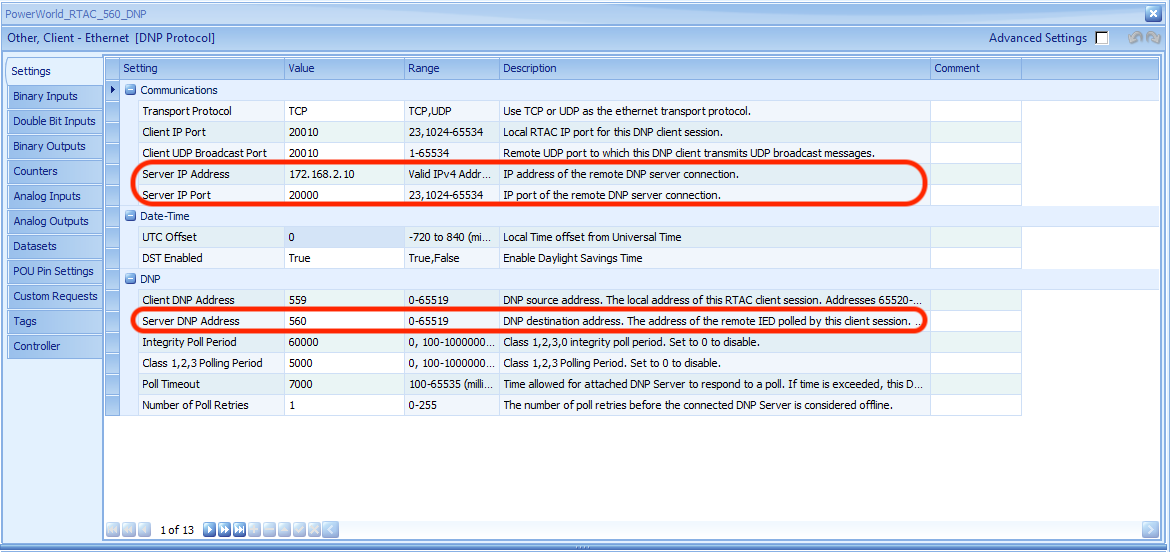}
  \caption{RTAC Configuration for Client PowerWorld\_RTAC\_560 for Substation 560.}
  \label{rtac}
\end{figure}

For RTAC, it is the DNP3 client for outstations in PWDS. To establish the DNP3 communication between PWDS and RTAC, RTAC's Ethernet port that connects to the host machine of PWDS and one of the host machine's Ethernet ports should be under the same LAN. In this paper, the connected RTAC Ethernet port's IP is 172.168.2.2 and one of the host machine's Ethernet port's IP is 172.168.2.10. Then, for each outstation, we can program a corresponding DNP3 client through SEL 5033 by inserting \textit{DNP Protocol} with \textit{Client-Ethernet} connection type. For clarity, we name the DNP3 client based on the substation ID. Within the client, the \textit{Server IP Address} is 172.168.2.10 and the \textit{Server IP Port} is 20000. The \textit{Server DNP Address} is the corresponding outstation number, which is the substation ID in this paper. As to \textit{Client IP Port} and \textit{Client DNP Address}, they can be configured based on user's preference as long as that port and address are not taken by other DNP3 applications/clients. The example of communication configuration for Substation 560 (GLEN ROSE) is shown in Figure~\ref{rtac}. For DNP3 communication settings, other parameters, such as \textit{Integrity Poll Period}, \textit{Class 1,2,3 Polling Period} and \textit{Poll Timeout}, are the default settings in SEL 5033. 

\begin{figure}[t]
\centering
\includegraphics[trim={0mm 0mm 0mm 0mm}, clip,height=2.3 in,width=3.5 in]{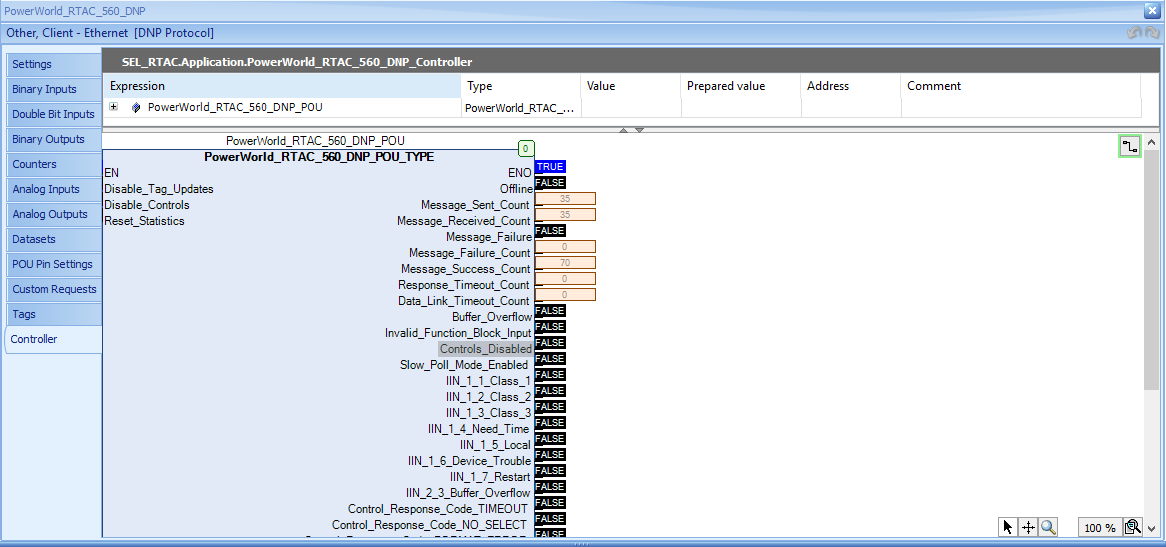}
  \caption{Client PowerWorld\_RTAC\_560 Controller}
  \label{rtac_check}
\end{figure}

\begin{figure}[t]
\centering
\includegraphics[trim={0mm 0mm 0mm 8mm}, clip,width=1\linewidth]{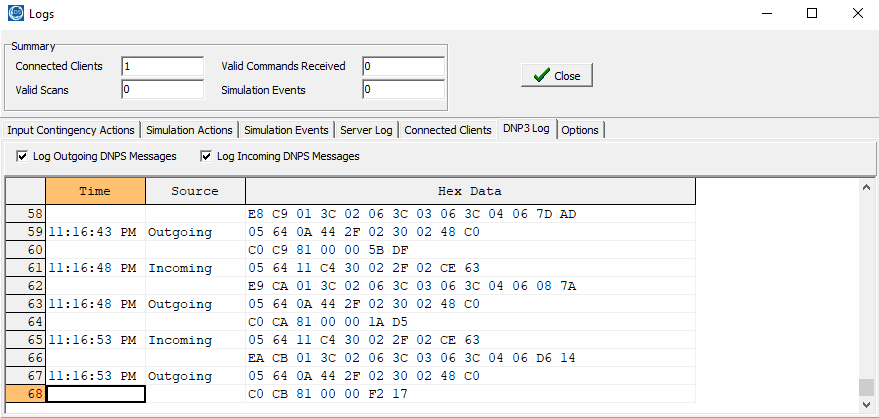}
  \caption{DNP3 Connection Logs} 
  \label{ds_check}
\end{figure}

After configuring the communication settings, we create \textit{Analog Input}, \textit{Analog Output}, \textit{Binary Input} and \textit{Binary Output} in RTAC to receive the data from PWDS. Once the configuration of RTAC is done, we can load the settings to RTAC through SEL 5033 by \textit{Go Online} option. Then, the RTAC configuration will be loaded to the hardware device for DNP3 communication. To check the communication between RTAC and PWDS, we can check the \textit{Controller} after the SEL 5033 is online. As shown in Figure~\ref{rtac_check}, a successful DNP3 connection's \textit{Offline} tag is \textbf{FALSE} and the \textit{Message\_Sent\_Count}, \textit{Message\_Received\_Count} and \textit{Message\_Success\_Count} are keeping increasing simultaneously. If there is any message fail to transmit, the \textit{Message\_Failure} will become \textbf{TRUE} and \textit{Message\_Failure\_Count} will show the number. From PWDS, we can check the \textit{Logs}, where shows the \textit{Connected Clients} and \textit{DNP3Log} as shown in Figure~\ref{ds_check}. There is an online period for SEL 5033. After the online period is passed, 
RTAC's settings are already configured and it can work as normal until the program has been updated and reloaded.

\subsection{Exemplar of DNP3 Reading and Control}
\label{example}

After the DNP3 communication between RTAC and PWDS is established, we can check the data in RTAC and send the control from RTAC to PWDS.

\begin{figure}[t!]
  \centering
  \subfloat[]{\includegraphics[trim={1mm 60mm 1mm 8mm}, clip,height=2.1 in,width=3.3 in]{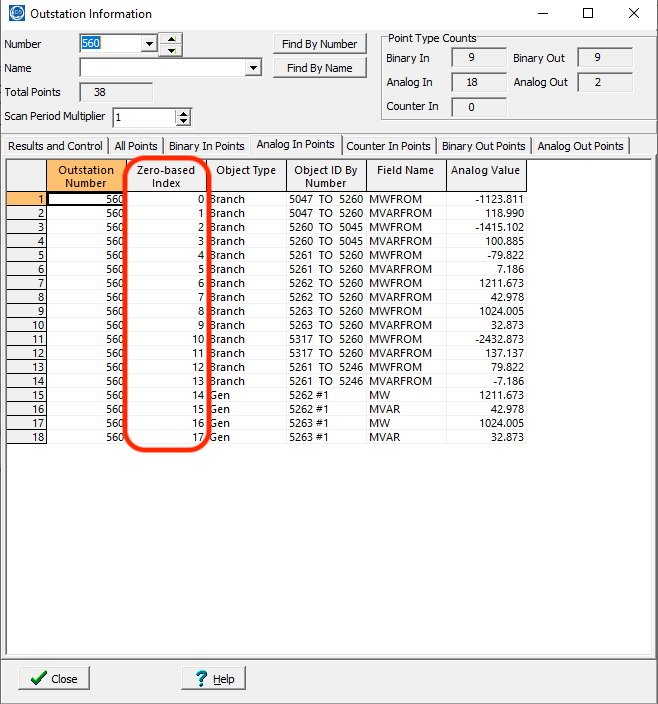}\label{DS_AI}}
  \vspace{-0.5cm}
  \subfloat[]{\includegraphics[trim={1mm 0mm 1mm 0mm}, clip,height=2.1 in,width=3.3 in]{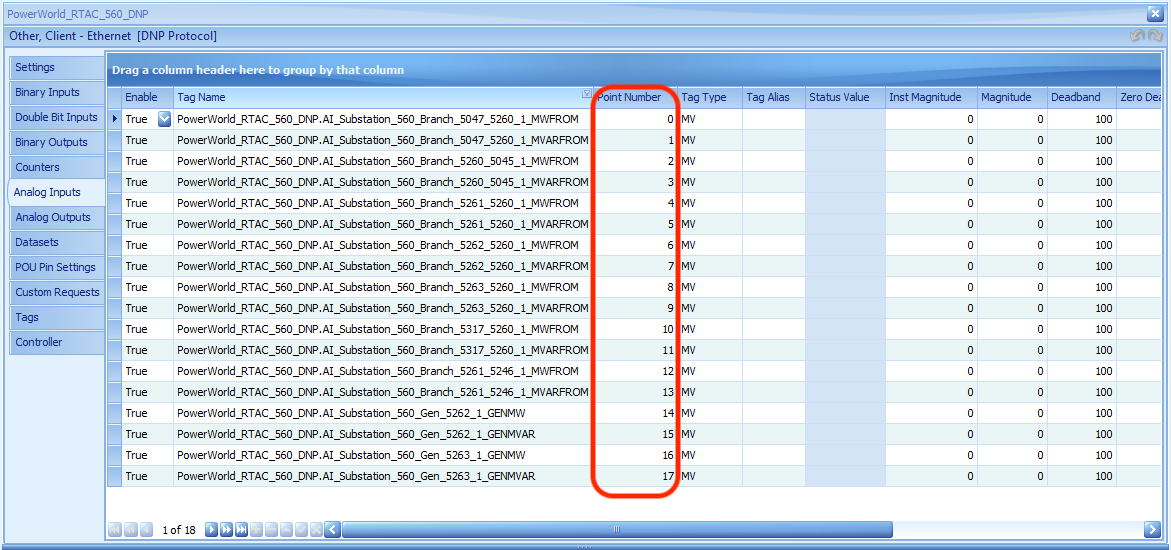}\label{RTAC_AI}}
\caption{DNP3 Analog Input Configuration in PWDS (Up) and RTAC (Bottom) for Data Mapping}\label{corresponding}
  \end{figure}

For simplicity and clarity, when we configure the PowerWorld case with \textit{DNP3Object}, we also create corresponding DNP3 tags for RTAC with the following pattern, \textit{DataType\_SubstationID\_DeviceType\_Keyfield\_DataName}. In this way, we can easily check the data from RTAC with detailed information of corresponding PowerWorld case information. For DNP3 data transition between client and server, it is based on the Zero-based Index (PWDS) and Point Number (RTAC) as shown in Figure \ref{corresponding} for \textit{Analog Input} data. RTAC collects the data from PWDS to corresponding tag based on the index and point number. This is the same for \textit{Analog Output}, \textit{Binary Input}, and \textit{Binary Output} data.

\begin{figure}[t]
\centering
\includegraphics[trim={0mm 30mm 0mm 0mm}, clip,height=1.8 in,width=3.3 in]{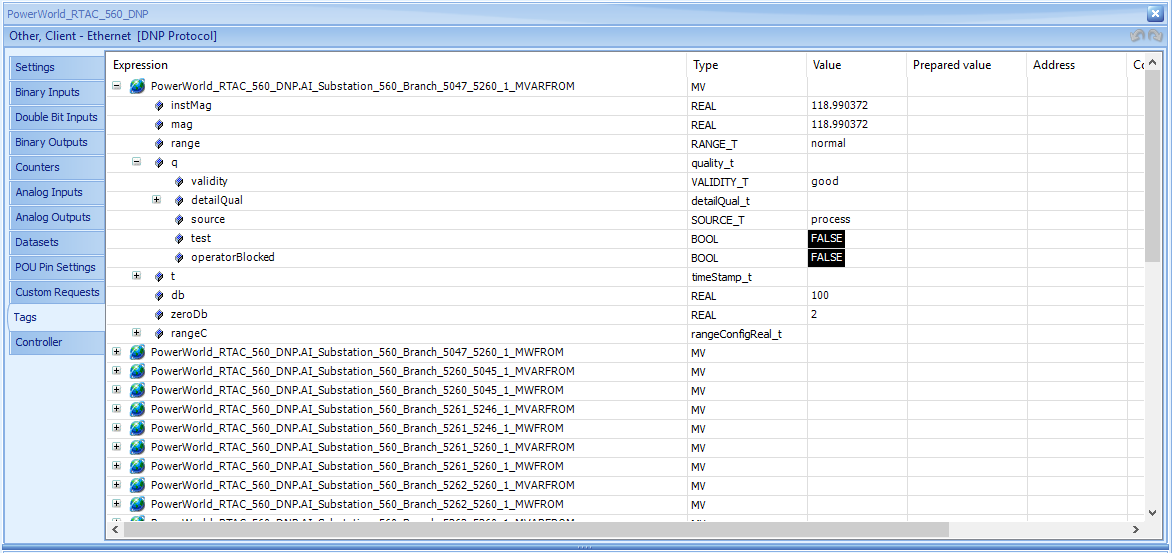}
  \caption{RTAC Analog Input Data for Branch 5047\_5260\_1 Reactive Power Flow.}
  \label{ai}
\end{figure}

\begin{figure}[t]
\centering
\includegraphics[trim={0mm 30mm 0mm 0mm}, clip,height=1.8 in,width=3.3 in]{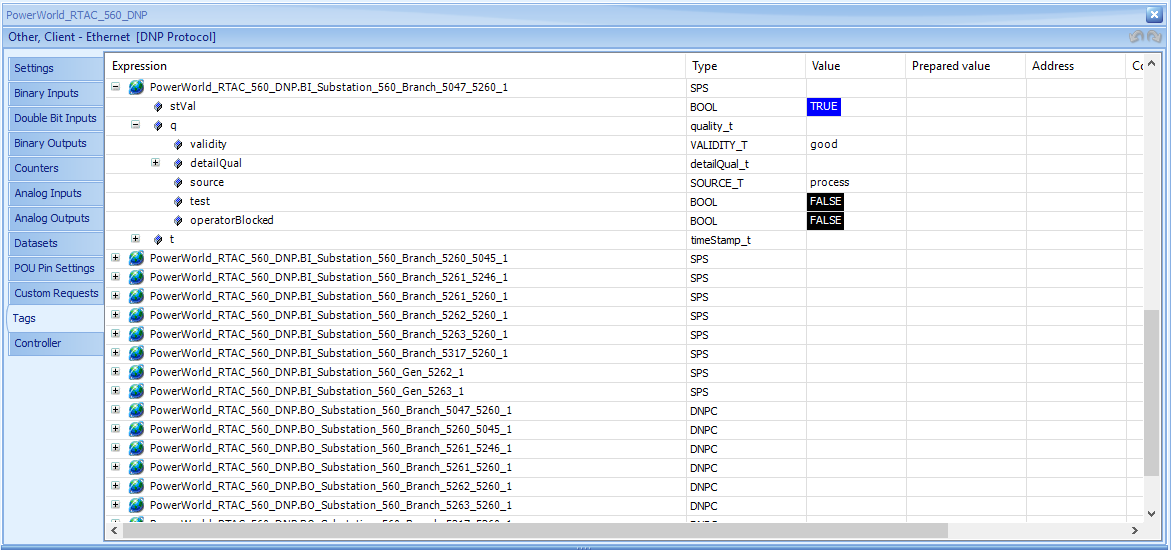}
  \caption{RTAC Analog Input Data for Branch 5047\_5260\_1 Status.}
  \label{bi}
\end{figure}

Once the DNP3 communication is successfully established, we can observe the data in RTAC \textit{Tag} list. As shown in Figure \ref{ai} and \ref{bi}, they show the \textit{Analog Input} and \textit{Binary Input} data for Branch 5047\_5260\_1's reactive power flow and status respectively and they are the same value as in PWDS. Besides the RTAC Controller tags, we can also check \textit{q} and its \textit{validity} value to see whether the DNP3 communication is successful or not. As shown in Figure \ref{ai} and \ref{bi}, both tags show \textbf{good}, so the current communication is established. When there is misconfiguration or cyber intrusion in the communication network, this tag will become \textbf{invalid}.

\begin{figure}[t!]
  \centering
  \subfloat[]{\includegraphics[trim={0mm 40mm 0mm 0mm}, clip,height=1.7 in,width=3.3 in]{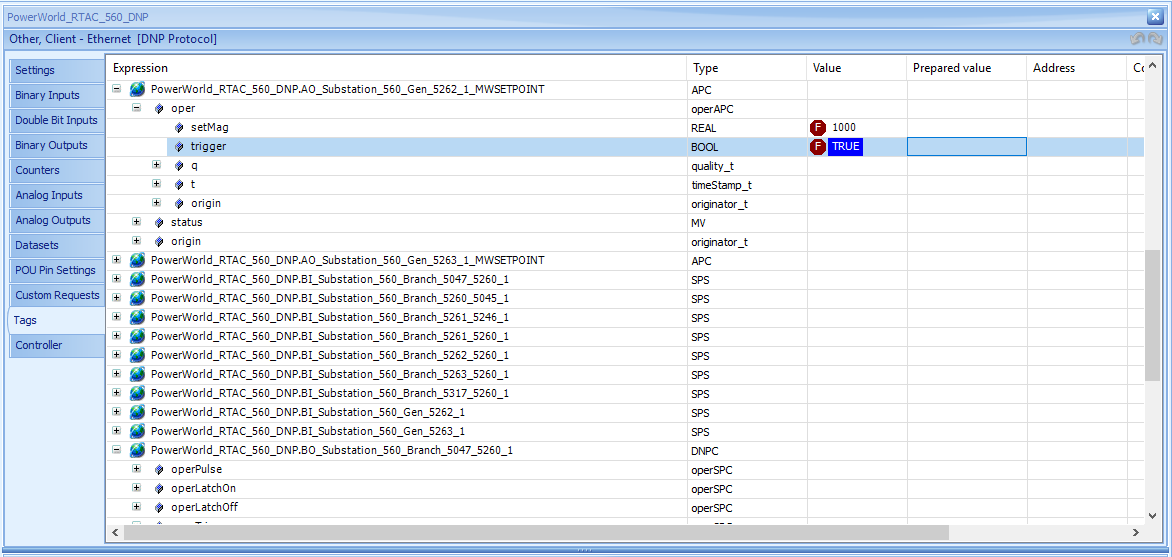}}
  \vspace{-0.5cm}
  \subfloat[]{\includegraphics[trim={0mm 40mm 0mm 0mm}, clip,height=1.7 in,width=3.3 in]{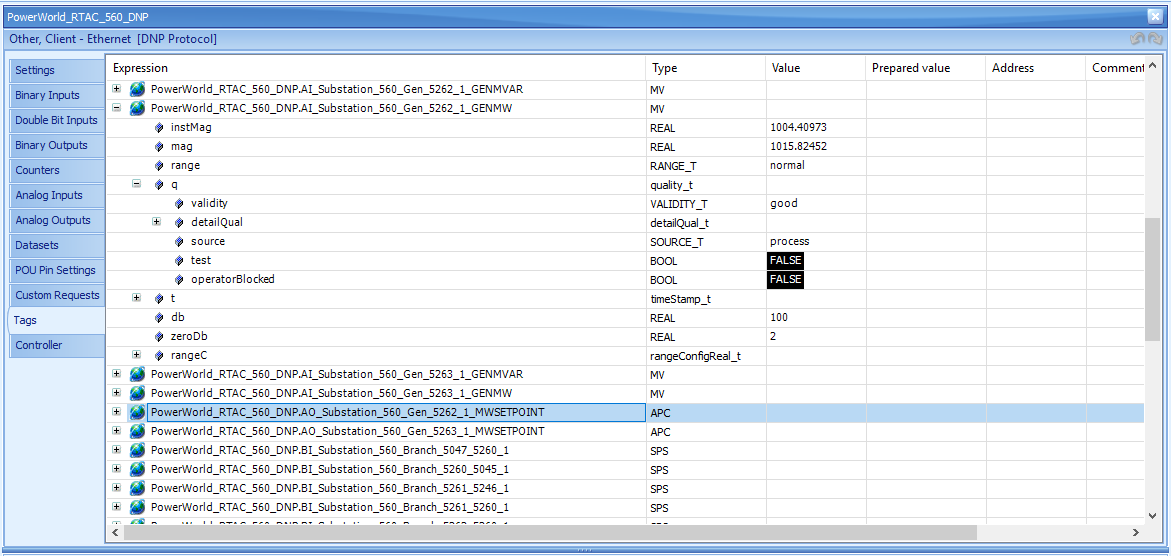}}
  
\caption{RTAC DNP3 Client Control Generator 5262\_1 real power output (Top) and the updated Analog Input reading (Bottom) }\label{ao}
  \end{figure}

RTAC client can also control the status of the device and change the generator \textit{MWSETPOINT} through \textit{Analog Output} and \textit{Binary Output} data. As shown in Figure \ref{ao}, we send a control command to Generator 5262\_1 in Substation 560 to change its real power output as 1000 MW through \textit{Analog Output} using the force value function in SEL 5033. After the generator receives the command, it gradually reduces its output from 1211 MW to 1004 MW. Because of the generator's exciter and governor model in PWDS, the generator's output will not reduce to 1004 MW immediately. In Figure \ref{ao}, there are two reading for Generator 5262\_1 real power output, one is \textit{instMag} whose value is 1004 MW and the other is \textit{mag} whose value is 1015 MW. The \textit{instMag} is the instantaneous value of corresponding tag's data, while the \textit{mag} is the value snapshot after \textit{instMag} exceeds the dead-band value, which is the time-stamped dead-banded event value \cite{sheet20193555}.

\begin{figure}[t!]
\centering
\subfloat[]{\includegraphics[trim={0mm 30mm 0mm 0mm}, clip,height=1.7 in,width=3.3 in]{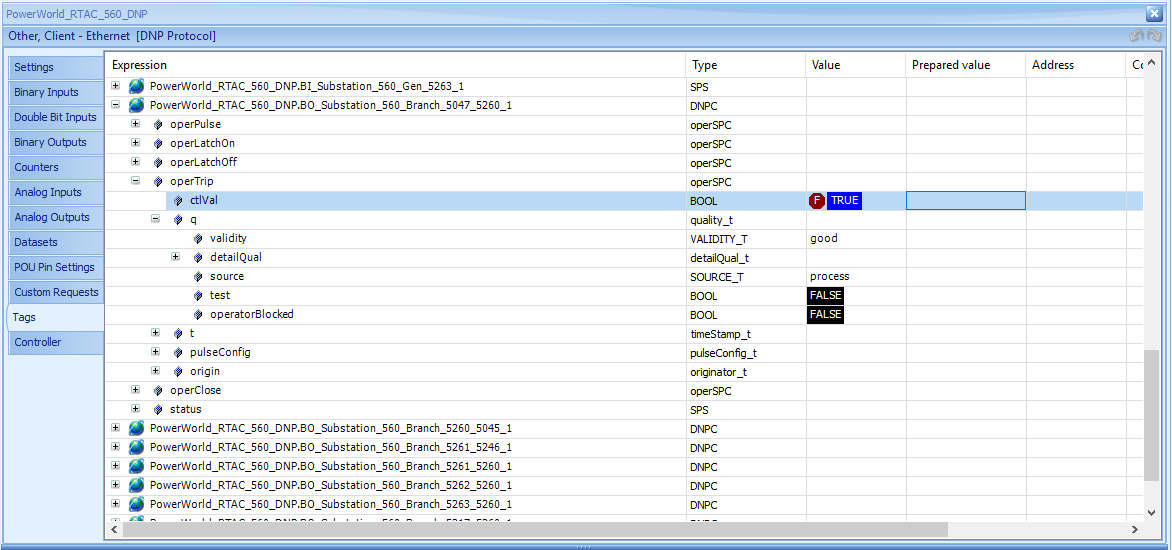}}
\vspace{-0.5cm}
\subfloat[]{\includegraphics[trim={0mm 30mm 0mm 0mm}, clip,height=1.7 in,width=3.3 in]{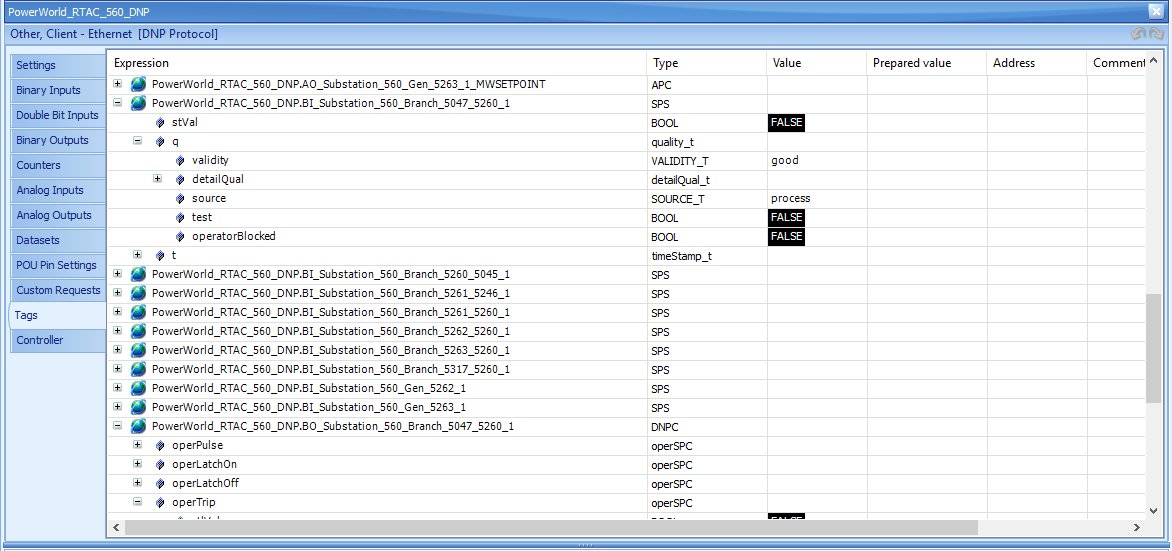}}
  
\caption{RTAC DNP3 Client Control Branch 5047\_5260\_1 to Open (Top) and the updated Binary Input reading (Bottom) }\label{bo}
  \end{figure}

Figure \ref{bo} shows the RTAC client open Branch 5047\_5260\_1 through Binary Output. After the command is executed, the branch will be open and updated \textit{Binary Input} for corresponding data will be \textbf{FALSE}.

\begin{figure}[t]
\centering
\includegraphics[trim={2mm 30mm 2mm 10mm}, clip,height=1.3 in,width=3.4 in]{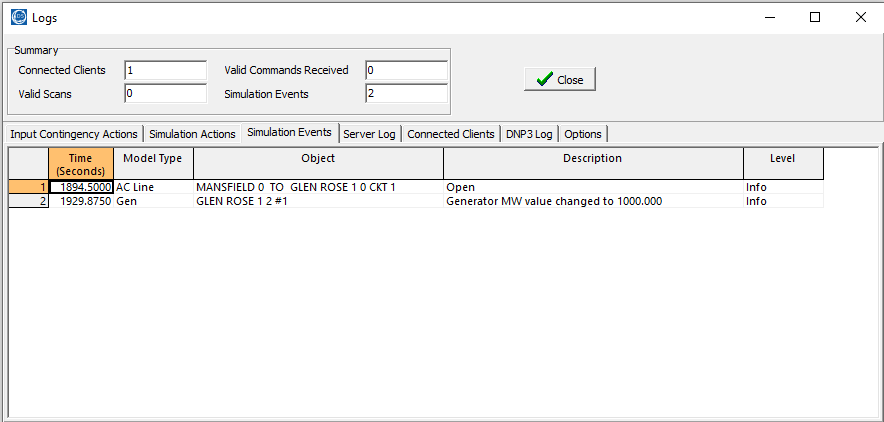}
  \caption{DS Logs for Operations from RTAC Client.}
  \label{event}
  
\end{figure}
All the commands sent from RTAC is logged in PWDS as shown in Figure \ref{event} with specific execution time and the counts of events.

\section{Conclusion}
\label{Conclusion}

In this paper, we present the cyber-physical testbed implementation of new functionality of PWDS that supports DNP3 communication capability, enabling real-time power system simulation in PWDS to generate DNP3 packets and deliver to DNP3 clients/masters. With an industrial automation and control device, this paper shows how to configure the synthetic power system case and RTAC to establish a successful DNP3 communication. It also shows the data that collected in RTAC \textit{Analog Input} and \textit{Binary Input} for corresponding measurement and status, and how the control command can be sent with \textit{Analog Output} and \textit{Binary Output} and committed in PWDS for real-time power system simulation.

The new functionality of DNP3 communication in PWDS provides a new mechanism to establish a hardware-in-the-loop testbed for power system cyber-physical security studies. PWDS can generate DNP3 packets based on configured DNP3 outstations and objects and deliver these packets to the DNP3 clients in industrial hardware, like RTAC, or EMS software, through a communication network. For future work, we can incorporate real or emulated communication network between PWDS and industrial hardware and software. Cyber intrusions can then be performed in the communication network, and the power system impacts can be observed in PWDS with real-time simulation; hardware devices can also detect such events with pre-defined alerts and control logic.

\section{Acknowledgement}
\label{Acknowledgement}
The work described in this paper was supported by funds from the US Department of Energy under award DE-OE0000895 and the National Science Foundation under Grant 1916142.

\bibliographystyle{IEEEtran}
\bibliography{sample}

\end{document}